\documentclass[preprint2]{aastex63}
\usepackage{amssymb }
\usepackage{amsmath}

\accepted{22 Aug 2021}

\submitjournal{AJ}

\begin{document}

\title{The Orbit of Planet Nine}

\correspondingauthor{Michael Brown}
\email{mbrown@caltech.edu}

\author[0000-0002-8255-0545]{Michael E. Brown}
\affiliation{Division of Geological and Planetary Sciences\\
California Institute of Technology\\
Pasadena, CA 91125, USA}

\author[0000-0002-7094-7908]{Konstantin Batygin}
\affiliation{Division of Geological and Planetary Sciences\\
California Institute of Technology\\
Pasadena, CA 91125, USA}

\begin{abstract}
The existence of a giant planet beyond Neptune --
referred to as Planet Nine (P9) --
 has been inferred from
the clustering of longitude of perihelion and
pole position of distant eccentric Kuiper belt objects (KBOs).
After updating calculations of observational
biases, we find that the
clustering remains significant at
the 99.6\% confidence level. We thus use these observations
to determine orbital elements of P9. A suite of numerical simulations 
shows that
the orbital distribution of the distant KBOs
is strongly influenced by the mass and orbital elements of
P9 and thus can be used to infer these parameters. 
Combining the biases with these 
numerical simulations, we calculate likelihood values for
discrete set of P9 parameters, which we then
use as input into a Gaussian Process emulator that allows a 
likelihood computation
for arbitrary values of all parameters. We use this emulator
in a Markov Chain Monte Carlo analysis to estimate
parameters of P9. 
We find
a P9 mass of $6.2^{+2.2}_{-1.3}$ Earth masses, semimajor axis of $380^{+140}_{-80}$ AU,
 inclination of $16\pm5^\circ$  and perihelion of $300^{+85}_{-60}$ AU.
Using samples of the orbital elements and estimates
of the radius and albedo of such a planet, we calculate the
probability distribution function of the on-sky position of
Planet Nine and of its brightness. 
For many reasonable assumptions,
Planet Nine is closer and brighter than initially expected, though the
probability distribution includes a long tail
to larger distances, and uncertainties in the radius and albedo
of Planet Nine could yield fainter objects.

\end{abstract}


\section{Introduction} \label{sec:intro}
Hints of the possibility of a massive planet well beyond the orbit
of Neptune have been emerging for nearly twenty years. The first
clues came from the discovery of a population
of distant eccentric Kuiper belt objects (KBOs) decoupled from interactions with Neptune \citep{2002Icar..157..269G,2003MNRAS.338..443E, 2006Icar..184..589G},
suggesting some sort of additional gravitational perturbation. 
While the first such decoupled objects were only marginally removed from
Neptune's influence and suggestions were later made that chaotic
diffusion could create similar orbits \citep{2017AJ....153..262B}, 
the discovery of 
Sedna, with a perihelion far removed from Neptune, clearly required
the presence
of a past or current
external perturber \citep{2004ApJ...617..645B}. 
Though the orbit of 
Sedna was widely believed
to be the product of perturbation by passing stars within the solar birth 
cluster \citep{2004AJ....128.2564M,2010ApJ...720.1691S, 2012Icar..217....1B}, 
the possibility of an external planetary perturber was also noted
\citep{2004ApJ...617..645B,2004AJ....128.2564M,2006Icar..184..589G}. More recently, \citet{2015Icar..258...37G} examined
the distribution of objects with very large semimajor axes but with 
perihelia inside of the planetary regime and concluded that their
overabundance can best be explained by the presence of an external planet
of mass $\sim$10 M$_e$ (where $M_e$ is the mass of the Earth) at a distance
of approximately 1000 AU. Simultaneously, \citet{2014Natur.507..471T} noted
that distant eccentric KBOs with semimajor axis $a>150$ AU all appeared
to come to perihelion approximately at the ecliptic and always travelling
from north-to-south (that is, the argument of perihelion, $\omega$, is 
clustered around zero), a situation that they speculated could be
caused by Kozai interactions with a giant planet, though detailed
modeling found no planetary configuration that could explain the observations.

These disparate observations were finally unified with the realization by
\citet{2016AJ....151...22B} that distant eccentric KBOs 
which are not under the
gravitational influence of Neptune are largely clustered 
in {\it longitude} of perihelion, meaning that their orbital axes are
approximately aligned, and simultaneously clustered in the orbital 
plane, meaning that their angular momentum vectors are approximately
aligned (that is, they share similar values of inclination, $i$, and 
longitude of ascending node, $\Omega$). 
Such a clustering is most simply explained by a giant
planet on an inclined eccentric orbit with its perihelion location
approximately 180 degrees removed from those of the clustered KBOs.
Such a giant planet would not only explain the alignment of the 
axes and orbital planes of the distant KBOs, but it would also naturally
explain the large perihelion distances of objects like Sedna, 
the overabundance of large semimajor axis low perihelion objects, 
the existence of a population of objects with orbits perpendicular
to the ecliptic,
and the apparent trend for distant KBOs to cluster about $\omega = 0$
(the clustering near $\omega=0$ is a coincidental consequence of the
fact that objects sharing the same orbital alignment and orbital plane
will naturally come to perihelion at approximately the same place
in their orbit and, in the current configuration of the outer
solar system, this location is approximately centered at $\omega \sim -40^\circ$).
The hypothesis that a giant planet on an inclined eccentric orbit
keeps the axes and planes of distant KBOs aligned was called the Planet Nine
hypothesis.

With one of the key lines of evidence for Planet Nine being the orbital
clustering, much emphasis has been placed on trying to assess whether or
not such clustering is statistically significant or could be a product 
of observational bias. In analyses of all available
contemporary data and their biases, 
\citet{2017AJ....154...65B} and \citet[hereafter BB19]{2019AJ....157...62B} find 
only a 0.2\% chance
that the orbits of the distant Kuiper belt objects (KBOs) are
consistent with a uniform distribution of objects. Thus the initial
indications of clustering from the original analysis
appear robust when an expanded data set that includes
observations taken over widely dispersed areas of the
sky are considered.
In contrast, \citet{2017AJ....154...50S}, \citet{2020PSJ.....1...28B}, and \citet{2021arXiv210205601N}  
using more limited -- and much more biased -- data sets, were unable to
distinguish between clustering and a uniform population. 
Such discrepant results are not surprising: 
BB19 showed that the data from
the highly biased OSSOS survey,
which only examined the sky in two distinct directions, 
do not have the sensitivity to detect the clustering already measured for
the full data set. \citet{2020PSJ.....1...28B} recognize that the sensitivity
limitations of the even-more-biased DES survey, which only examined the sky in a single
direction, precluded them from being able to constrain clustering.
It appears that Napier et al., whose data set is dominated by the combination of the
highly-biased OSSOS and DES surveys, suffers
from similar lack of sensitivity, though Napier et al. do not provide
sensitivity calculations that would allow this conclusion to be confirmed. 
Below, we update the calculations of BB19 and demonstrate
that the additional data now available continues to support the statistical
significance of the clustering. We thus continue to suggest that
the Planet Nine hypothesis remains the most viable explanation for
the variety of anomolous behaviour seen in the outer solar system,
and we work towards determining orbital parameters of Planet Nine.

Shortly after the introduction of the Planet Nine
hypothesis, attempts were
made to constrain various of the orbital elements of the planet.
\citet{2016ApJ...824L..23B} compared the observations to some early simulations of
the effects of Planet Nine on the outer solar system and showed
that the data were consistent with a Planet Nine with a mass between
5 and 20 Earth masses, a semimajor axis between 380 and 980 AU,
and a perihelion distance between 150 and 350 AU. 
Others sought to use the possibility that the observed objects
were in resonances to determine parameters \citep{2016ApJ...824L..22M}, though
\citet{2018AJ....156...74B} eventually showed that this route is not feasible.
\citet{2017AJ....153...91M} invoked simple metrics to compare simulations and
observations, and \citet{2019PhR...805....1B} developed a series 
of heuristic metrics to 
compare to a large suite of simulations and provided the best
constraints on the orbital elements of Planet Nine to date. 

Two problems plague all of these attempts at deriving parameters of 
Planet Nine. First, the metrics used to compare models and observations,
while potentially useful in a general sense,
are {\it ad hoc} and difficult to justify statistically. As importantly,
none of these previous methods has attempted to take into account
the observational biases of the data.
While we will demonstrate here that the clustering of orbital parameters in
the distant Kuiper belt is unlikely a product of observational bias, observational bias {\it does} affect the orbital
distribution of distant KBOs which have been discovered. Ignoring these
effects can potentially bias any attempts to discern orbital properties of
Planet Nine.

Here, we perform the first rigorous statistical assessment of the orbital
elements of Planet Nine. We use a large suite of 
Planet Nine simulations,
the observed orbital elements of the distant Kuiper belt, as well as
the observational biases in their discoveries, to develop a detailed
likelihood model to compare the observations and simulations. 
Combining the likelihood models from all of the simulations, we
calculate probability density functions for all orbital parameters
as well as their correlations, providing a map to 
aid in the search for 
Planet Nine.

\section{Data selection}
The existence of a massive, inclined, and eccentric planet  
beyond $\sim$250 AU has been shown to be able to
cause multiple dynamical effects, notably including a clustering of longitude of perihelion, $\varpi$, and of pole position (a combination of 
longitude of ascending node, $\Omega$, and inclination, $i$) for 
distant eccentric KBOs. 
Critically, this clustering is only
strong in sufficiently distant objects whose
orbits are not strongly affected by interactions with Neptune \citep{2016AJ....151...22B, 2019PhR...805....1B}. Objects 
with perihelia closer to the semimajor
axis of Neptune, in what is sometimes referred to as the
``scattering disk,'' for example, have the strong clustering effects of
Planet Nine disrupted and are more uniformly situated \citep[i.e.][]{2017AJ....153...33L}. 
In order to not dilute the effects of Planet Nine with random
scattering caused by Neptune, we thus follow the original
formulation of the Planet Nine hypothesis and 
restrict our analysis to
only the population not interacting with Neptune. 
In  \citet{2019PhR...805....1B}
we use numerical integration to 
examine the orbital history of each known distant object and 
classify them as stable, meta-stable, or unstable, based
on the speed of their semimajor axis diffusion. In that analysis,
all objects with $q<42$ AU are unstable with respect to perihelion
diffusion, while all objects with $q>42$ AU are stable or meta-stable.
Interestingly, 11 of the 12 known KBOs with $a>150$ AU and $q>42$ AU
have longitude of perihelion clustered between $7<\varpi<118^{\circ}$,
while only 8 of 21 with $30<q<42$ AU are clustered in this region,
consistent with the expectations from the Planet Nine hypothesis.
We thus settle on selecting all objects at $a>150$ AU
with perihelion distance, $q>42$ AU for analysis for both the
data and for the simulations below.

A second phenomenon could also dilute the clustering
caused by Planet Nine. Objects which are scattered {\it inward}
from the inner Oort cloud also appear less clustered than the longer-term
stable objects \citep{2021ApJ...910L..20B}. 
These objects are more difficult to exclude
with a simple metric than the Neptune-scattered objects, though excluding
objects with extreme semimajor axes could be a profitable approach. 
Adopting our philosophy from the previous section, we exclude the
one known object in the sample with $a>1000$ AU as possible contamination
from the inner Oort cloud. While we again cannot know for sure if this
object is indeed from the inner Oort cloud, removing the object can only
have the effect of decreasing our sample size and thus increasing the uncertainties
in our final orbit determination, for the potential gain of decreasing any biases
in our final results. 

The sample with which we will compare our observations thus includes
all known multi-opposition KBOs with $150<a<1000$ AU and $q>42$ AU reported as 
of 20 August 2021. Even after half of a decade of intensive search for distant objects in
the Kuiper belt, only 11 fit this stringent criteria for comparison
with models. The observed orbital elements of these 11 are shown in Table 1.
These objects are strongly clustered in $\varpi$ and pole position,
though observational biases certainly can affect this observed
clustering.

\begin{figure*}[t]
\plotone{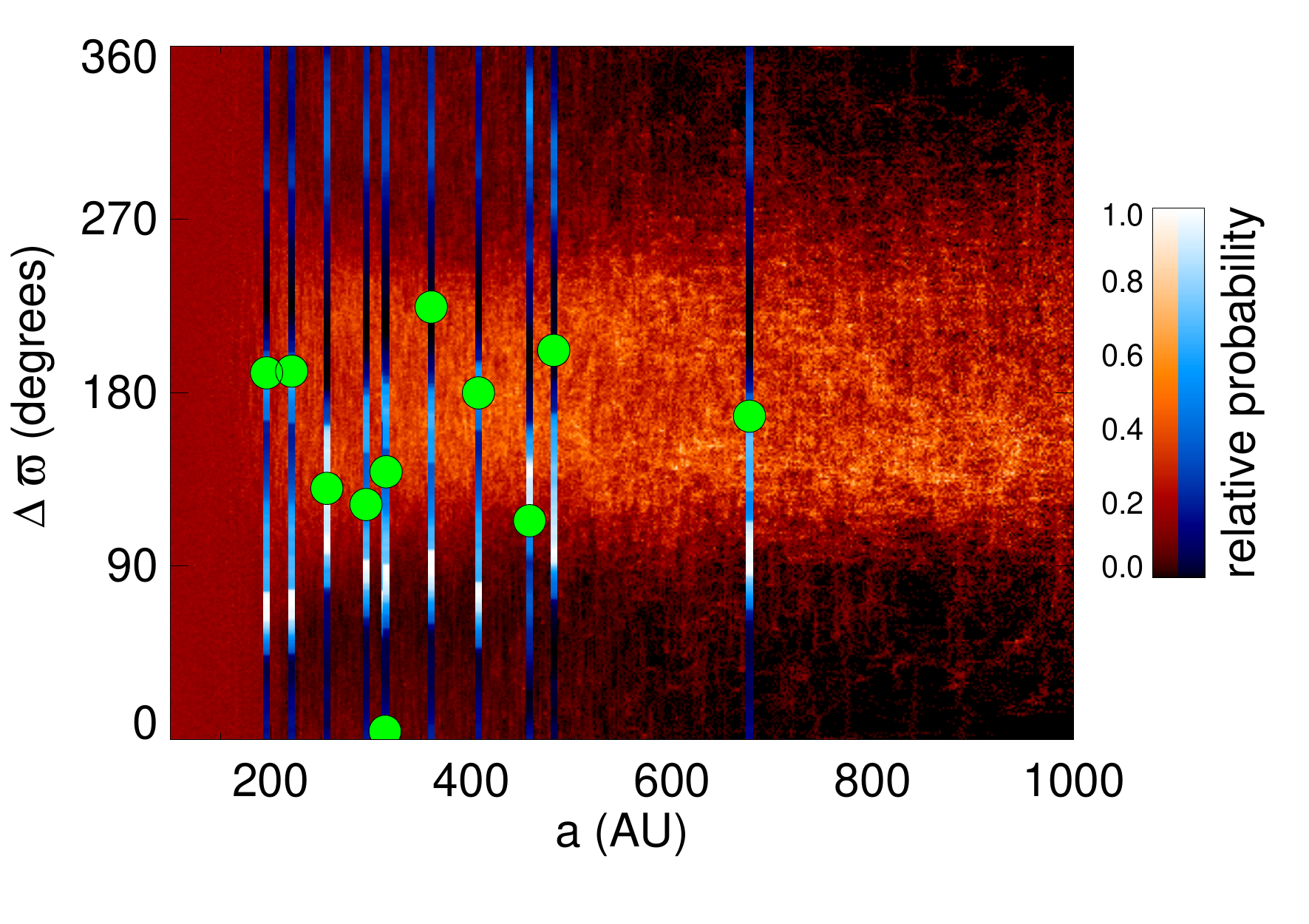}
\caption{Semimajor axes versus $\varpi$ for the 11 KBOs
of our sample (green points). The points are plotted as
$\Delta \varpi$, defined as $\varpi-\varpi_9$, where here we plot the 
points for an assumed value of $\varpi_9=254^\circ$.
For each known distant KBO we show a one-dimensional projection
of the bias with respect to $\varpi$ (blue). While consistent bias
exists, the $\varpi$ cluster is approximately $90^\circ$ removed from
the direction of bias.
We also show the probability density of $\varpi$ versus
semimajor axis in the 
maximum likelihood model with $m_9=5$ M$_e$, $a_9=300$ AU, $e_9=0.15$
and $i_9=16^\circ$ (red).
The density plot is normalized at every semimajor axis to better
show the longitudinal structure. Note that this comparison is simply for
visualization; the full maximum-likelihood model compares the full
set of orbit elements of each object to the simulations and also
incorporates the observational biases on each observed objects.}
\end{figure*}
\begin{figure*}[ht!]
\includegraphics[scale=1.1]{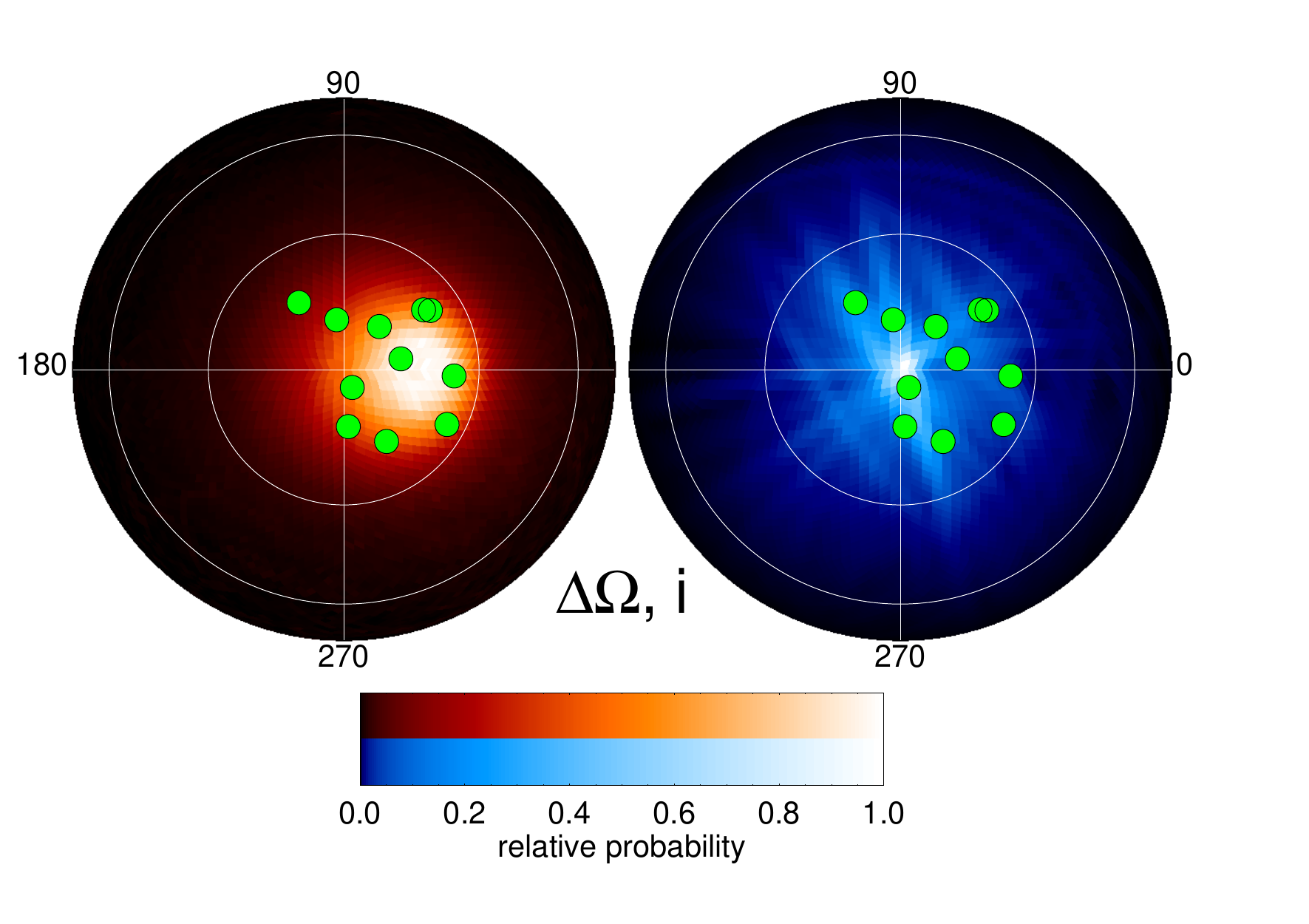}
\caption{A comparison of the projection of the pole position 
of the distant detached
KBOs (the green points show ($\sin i \cos \Delta \Omega$, $\sin i \sin \Delta \Omega$), where $\Delta \Omega$ is the difference between the longitude of
ascending node of the observed object and of the modeled Planet Nine, assumed 
to be $108^\circ$ here) and a density plot of their expected values
in the maximum likelihood model (red background). In blue we show an average
of the two-dimensional projection of
the pole position bias of all of the objects. While strong bias in pole
position exists, no preferential direction is apparent. White circles
indicate 30 and 60 degree inclinations.
}
\end{figure*}

\begin{deluxetable}{cccccc}
\tablecaption{Orbital elements of all reported objects with $150<a<1000$ and $q>42$ AU.}

\tablehead{
    \colhead{name} & \colhead{a} & \colhead{e} & \colhead{i} & \colhead{$\Omega$} & \colhead{$\varpi$} \\
 \colhead{} & \colhead{AU} & & \colhead{deg.} & \colhead{deg.} & 
\colhead{deg.}}
    \startdata
       2000CR105&218&0.80&22.8&128.3& 85.0\\
2003VB12&479&0.84&11.9&144.3& 95.8\\
2004VN112&319&0.85&25.6& 66.0& 32.8\\
2010GB174&351&0.86&21.6&130.8&118.2\\
2012VP113&258&0.69&24.1& 90.7& 24.2\\
2013FT28&312&0.86&17.3&217.8&258.3\\
2013RA109&458&0.90&12.4&104.7&  7.5\\
2013SY99&694&0.93& 4.2& 29.5& 61.6\\
2013UT15&197&0.78&10.7&192.0& 84.1\\
2014SR349&302&0.84&18.0& 34.8& 15.7\\
2015RX245&412&0.89&12.1&  8.6& 73.7\\
\enddata
\tablecomments{As of 20 August 2021.}
\end{deluxetable}

\section{Bias}
All telescopic surveys contain observational biases. 
Correctly understanding and implementing these biases into
our modeling is critical to correctly using the observations to extract
orbital parameters of Planet Nine.
 BB19 developed a method to use the
 ensemble of all 
known KBO detections to estimate a full geometric observational
bias for individual distant KBOs. For each of
the distant KBOs, they create the 
function
\begin{equation}
B_j^{(a,e,H)_j}[(i,\varpi,\Omega)|U],
\end{equation}
where, for our case, $j$ represents one of the 11 distant KBOs of the sample 
and $B_j^{(a,e,H)_j}$ is the probability that 
distant KBO $j$, with semimajor axis, eccentricity, and absolute magnitude $(a,e,H)_j$
would be detected with orbital angles $i$, $\varpi$, and $\Omega$, if the population were
uniformly distributed in the sky, given $U$, the
ensemble of all known KBO detections. 
The details
of the method are explained in BB19,
but, in short, it relies on the insight that every detection
of every KBO 
can be thought of (with appropriate caveats)
as an observation at that position in the sky
that {\it could have} detected an equivalent
object $j$ with $(a,e,H)_j$ if, given the required
orbital angles $(i, \varpi, \Omega)$ to put object $j$
at that position in the sky, the object would be predicted
to be as bright as or brighter at that sky position
than the detected KBO. 
For each sample object $j$ with $(a,e,H)_j$, 
the ensemble of all KBO detections
can thus be used to estimate all of the orbital angles
at which the object could have been detected. This 
collection of orbital angles at which an object with $(a,e,H)_j$
could have been detected represents the bias in $(i, \varpi, \Omega)$
for object $j$. While biases calculated with this method
are strictly discrete, we smooth to one degree resolution in all
parameters for later application to our dynamical simulations.

Note that this method differs from
bias calculations using full survey simulators. 
It does not rely on knowledge of the survey details
of the detections, but rather just the fact of the detection itself. 
Comparison of these bias calculations with the bias calculated
from a full survey similar for the
OSSOS survey shows comparable results (BB19).

Of the objects in our sample,
all were included in the BB19 
calculations with the exception of
2013 RA109, which had not been announced at the time of the original
publication. We reproduce the algorithm of BB19 to calculate a bias probability function for this object.

While the bias is a separate 3-dimensional function for each object, 
we attempt to give
an approximate visual representation of these biases in Figure 1, 
which collapses the bias of each object in $\varpi$ into a single
dimension.
As can be seen, a strong observational bias in $\varpi$ exists, but the observed clustering
is approximately $90^\circ$ removed from the position of this bias.
Figure 2 shows the bias in pole position. While, again, each object
has an individual bias, the 
pole position biases are sufficiently 
similar that we simply show the sum of all of the biases, collapsed
to two dimensions. Strong pole position biases exist, but none
which appear capable of preferentially biasing the pole in any
particular direction.

With the bias function now available, we re-examine the 
statistical significance of the angular clustering
of the distant KBOs by updating the analysis of BB19 
for the objects in our current analysis set. 
As in that analysis, we perform $10^6$ iterations
where we randomly chose $(i,\varpi,\Omega)$ for the 11 objects of our sample
assuming uniform distributions in $\varpi$ and $\Omega$ and
a $\sin i \exp(-i^2/2\sigma^2)$ distribution with $\sigma=16^\circ$ 
for $i$, and project these to the four-dimensional space of
the canonical Poincar\'e variables $(x,y,p,q)$, corresponding roughly to
longitude of perihelion ($x,y$) and pole position ($p,q$)
(see BB19 for details). For each of the iterations we compute the average
four-dimensional position of the 11 simulated sample objects 
and note whether or not
this average position is more distant than the average position of
the real sample. This analysis finds that the real data are more extreme
than 99.6\% of the simulated data, suggesting only a 0.4\% chance that these
data are selected from a random sample. 
Examination of Figures 1 and 2 give a good 
visual impression of why this probability is so low. The data are distributed
very differently from the overall bias,
contrary to expectations for a uniform sample.

The significance of the clustering retrieved here is slightly worse
than that calculated by
BB19. While one new  distant object has been added to the sample, the
main reason for the change in the significance is that, after the
\citet{2019PhR...805....1B} analysis, we now understand much better which
objects should most be expected to be clustered by Planet Nine, thus
our total number of objects in our sample is smaller. Though this
smaller sample leads to a slightly lower clustering significance,
we nonetheless recommend the
choice of $150<a<1000$ AU and $q>42$ AU for any analyses going forward
including newly discovered objects.

With the reassurance that the clustering is indeed robust, 
we now turn to using the biases to
help determine the orbital parameters of Planet Nine. 

\section{Planet Nine orbital parameter estimation}
To estimate orbital parameters of Planet Nine, we require a likelihood
model for a set of orbital parameters given the data on the observed distant
KBOs. In practice, because of the structure
of our bias calculations, which only account for on-sky geometric biases and do not attempt to explore biases in semimajor axis or perihelion, 
we reformulate this likelihood
to be that of finding the observed value of ($i,\varpi,\Omega)_j$
given the specific value of $(a,e)_j$ for each distant object $j$.
Conceptually, this can be thought of as calculating the probability
that an object with a measured value of $a$ and a measured value of 
$q$ would be found to have the measured values of $i$, 
$\varpi$, and $\Omega$ for a given set of Planet Nine orbital parameters.

The random variables for this model  
are the mass of the planet, $m_9$, the semimajor axis, $a_9$, the eccentricity, $e_9$, the inclination, $i_9$, the longitude of perihelion, $\varpi_9$,
and the longitude of ascending node, $\Omega_9$. As the effects of Planet Nine are now understood to be mainly secular \citep{2016A&A...590L...2B}, the position of Planet Nine within 
its orbit (the mean anomaly, $M_9$) does not affect the outcome, so it
is unused. We thus write the likelihood function of the $j^{\rm th}$ KBO
in our data set as:
\begin{equation}
\mathcal{L}_j^{(a,e)_j}[(m_9,a_9,e_9,i_9,\varpi_9,\Omega_9)|(i,\varpi, \Omega)_j],
\end{equation}
where the $(a,e)_j$ superscript refers to the fixed value of $a$
and $e$ for object $j$. The full likelihood, $\mathcal{L}_{\rm P9}$ is  the
product of the individual object likelihoods.
The likelihood
of observing $(i,\varpi,\Omega)_j$ given a set of Planet Nine parameters depends on
both the physics of Planet Nine and the observational biases.

\subsection{Simulations}
While $\mathcal{L}_{\rm P9}$ is presumably a continuous function
of the orbital parameters,
we must calculate the value at discrete locations using 
numerical simulations.
We perform 121 of these simulations at manually chosen values of 
$m_9$, $a_9$, $e_9$, and $i_9$, as detailed below. 
The two angular parameters, $\varpi_9$ and $\Omega_9$, 
yield results that are rotationally symmetric so we need not 
individually simulate these results but rather rotate our
reference frame to vary these parameters later. We
 set $M_9=0$ for the starting position of all simulations as this
 parameters does not affect the final orbital distributions.

To save computational time, previous Planet Nine simulations have often
included only the effects of Neptune plus a $J_2$ term to simulate the
combined orbit-averaged torque of the three inner gas giants. 
While this approach 
captures the relevant processes at the qualitative level, 
here, as we are interested
in a detailed comparison with observations, we fully include all
four inner giant planets. 
For each independent simulation 
a set of between 16,800 and 64,900 
test particles is initially distributed with
semimajor axis between 150 and 500 AU, perihelion between 30 and 50 AU,
inclination between 0 and $25^\circ$, 
and all other orbital angles randomly distributed. The orbits of
the 5 giant planets and test particles are integrated using the 
mercury6 gravitational dynamics software package \citep{1999MNRAS.304..793C}.
To carry out the integrations we
used the hybrid symplectic/Bulirsch-Stoer algorithm of the package,
using a  time step of 300 days which is adaptively reduced to
directly resolve close encounters.
Objects that collide with planets or reach
$r<4.5$ or $r>10000$ AU are removed from the simulation for 
convenience. The orbital
elements of all objects are defined in a plane in which 
the angles are referenced to the initial plane of the four interior 
giant planets. 
As Planet Nine precesses the plane of the planets, however, 
the fixed reference coordinate system no longer corresponds to the 
plane of the planets. Thus, after the simulations are completed, we recompute
the time series of ecliptic-referenced angles 
by simply rotating to a coordinate system aligned
with the orbital pole of Jupiter. In this rotation we keep the longitude
zero-point  fixed so that nodal precession of test particles
and Planet Nine can be tracked.

A total of 121 simulations was performed, varying 
the mass ($m_9$), semimajor axis ($a_9$), eccentricity ($e_9$),
and inclination ($i_9$). 
Parameters for Planet Nine were chosen by hand in an attempt to 
explore a wide range of parameter space and find the 
region of maximum likelihood. The full set of parameters explored
can be seen in Table~2. Examination of 
the initial results from
these simulations confirms the conclusions of \cite{2019PhR...805....1B}: varying
the orbital parameters of Planet Nine produces large effects on
the distant Kuiper belt (Fig. 3). We see, for example, that
fixing all parameters but increasing $m_9$ smoothly narrows the
spread of the distant cluster (the feature labeled ``cluster width'' in
Figure 3). Increasing $i_9$ smoothly moves
the orbital plane of the clustered objects to follow the orbital
plane of Planet Nine, until, at a values above $i_9~>30^{\circ}$, the increased
inclination of Planet Nine tends to break the clustering entirely (Figure 4). 
Increasing $m_9$ also leads to a decrease in the distance to the transition
between unclustered and clustered objects (the feature labeled the ``wall'' 
in Figure 3), while increasing the perihelion distance of Planet Nine ($q_9$)
increases the distance to the wall. Many other more subtle effects
can be seen in the full data set. While we point out all of
these phenomena, our point is not to parameterize or make use of any of 
them, but rather to make the simple case that the specific orbital
parameters of Planet Nine cause measurable effects on the distributions
of objects in the distant Kuiper belt. Thus, we should be able to use
the measured distributions to extract information about the orbital 
parameters of Planet Nine. We will accomplish this task through our
full likelihood model.
\begin{figure*}[ht!]
\epsscale{1.}
\plotone{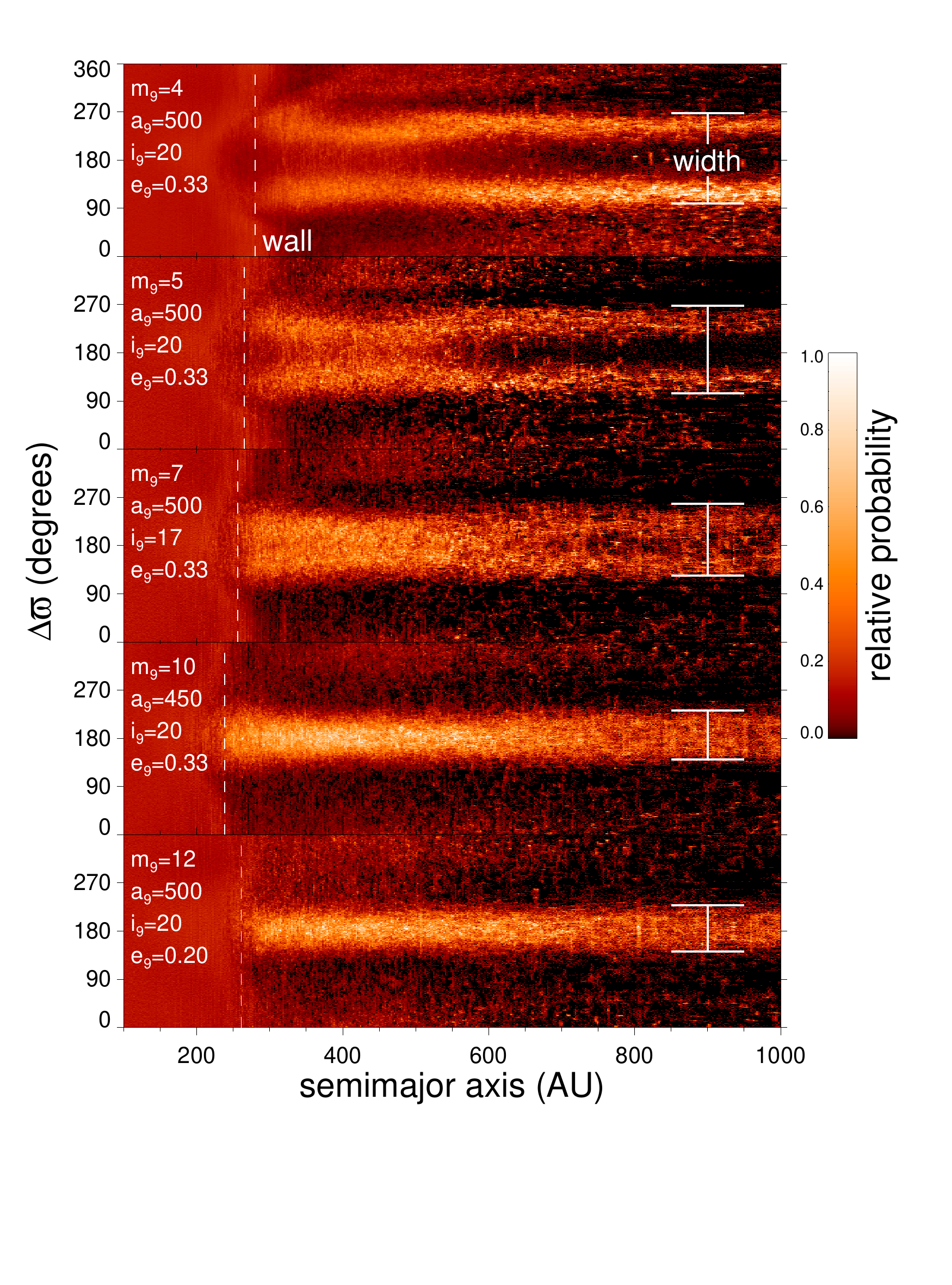}
\caption{Probability density of $a$ versus $\Delta \varpi$ for a
variety of simulated Planets Nine, in the same format as Fig.1. 
At the lowest masses the cluster appears double-peaked as the clustered
objects librate and spend greater amounts of time at their inflection
points.
The dashed line labeled ''wall'' shows the transition between 
the nearby uniform population and the more distant clustered population.
This transition distance decrease with increasing $m_9$ and decreasing
$a_9$ and $q_9$. The width of the cluster decreases with increasing 
$m_9$. Systematic changes such as these demonstrate that the orbital
distribution of the distant KBOs is strongly influenced by the
orbital parameters of Planet Nine.}
\end{figure*}
\begin{figure*}
\plotone{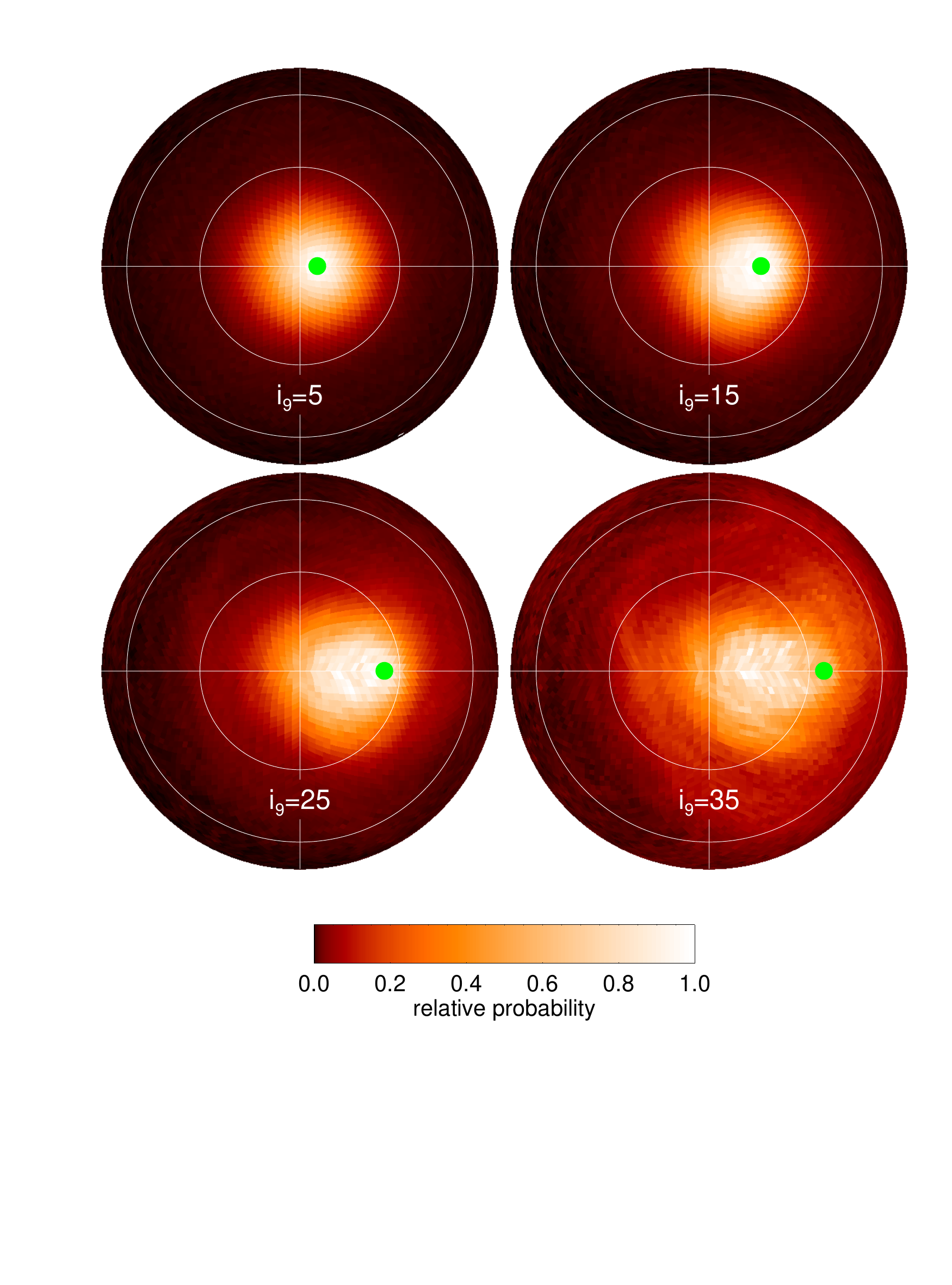}
\caption{The probability distribution of the pole positions of
KBOs with $a>150$ AU and $q>42$ AU for four Planet Nine models. 
All models have $m_9=7$ M$_{\rm earth}$, $a_9=500$ AU, and $e_9=0.33$
and the effect of changing $i_9$ can be seen. The green dot shows the 
pole position of Planet Nine in the simulations. The white circle show
30 and 60$^{\circ}$ inclinations. For small inclinations the 
distant KBOs track the inclination of Planet Nine. As $i_9$ increases,
however, the overall longitudinal clustering diminishes and the pole
positions cluster less tightly around that of Planet Nine.}
\end{figure*}

\subsection{Kernel density estimation}

Each numerical simulation contains snapshots of the orbital distribution of the outer solar system for a finite number
of particles. We use kernel density estimation to estimate
a continuous function for the probability distribution
function (PDF) from these discrete results of each 
simulation, that is, we seek the probability of
observing an object at $(i,\varpi,\Omega)_j$ given $(a,e)_j$ for
each simulation.
The early times of the simulations
contain a transient state that appears to reach something like a steady-state
in orbital distribution after $\sim 1$ Gyr. We thus discard these initial
time steps and only include the final 3 Gyr in our analysis. In all
simulations the number of surviving objects continues to decrease 
with time, with a wide range in variation of the ejection rate that depends
most strongly on P9 mass and perihelion distance.

For each numerical model, $k$, and each observed KBO, $j$, 
we repeat the following steps. First, we collect all modeled objects 
that pass within a defined
smoothing range of $a_j$ and $q_j$, the parameters of the 
observed KBO. Because of our finite number of particles, smoothing
is required to overcome the shot noise which would otherwise
dominate the results. Based on our observation that the behaviour of the modeled
KBOs changes rapidly with changing semimajor axis around the transition
region (we do not know this transition region {\it a priori} but it is
within 200-400 AU in all the simulations; see Figures 1 and 3, for example)
but changes little at large semimajor axes, 
we define the smoothing
range in $a$ as a constant value of 5\% for $a_j<230$ AU, but, because 
the number of particles in the simulations declines with increasing semimajor
axis, we allow the smoothing distance to
linearly rise to 30\% by $a_j=730$ AU. 
For perihelia beyond 42 AU, we observe little change in behaviour
as a function of $q$, so we define a simple smoothing length 
of $q_j \pm 10$ AU with a lower 
limit of 42 AU (which is the limit we imposed on the observed KBOs). 
The main effect of these two smoothing parameters
will be to slightly soften the sharp transition region (``the wall'')
which, in practice, will contribute to the uncertainties in our
derived mass, eccentricity, and, semimajor axis.

We select 
all of the modeled KBOs at times after the initial Gyr of simulation
that pass within these $a$ and $q$ limits at any time step, and we
weight them with two Gaussian kernels, each with a $\sigma$ 
equal to 
half of the smoothing distances defined above. 
The selected objects now all contain similar 
semimajor axis and perihelion distance as the $j^{\rm th}$ observed 
KBO, and their normalized distribution gives the probability that such
an observed KBO would have a given inclination, longitude of perihelion, and longitude of ascending node. 
At this point the simulated values of $\varpi$ and $\Omega$ are
all relative to $\varpi_9$ and $\Omega_9$, rather than in
an absolute coordinate system. We refer to these relative
values as $\Delta\varpi$ and $\Delta\Omega$.

We create the three-dimension probability distribution function
of $(i, \Delta\varpi, \Delta\Omega)$ by selecting a value
of $\Delta\varpi$ and then constructing a probability-distribution
function of the pole position $(\sin i\cos \Omega, \sin i\sin\Omega)$,
again using kernel density estimation now 
using a Gaussian kernel with $\sigma=2^{\circ}$ in great-circle 
distance from the pole position
and $\sigma=10^{\circ}$ in longitudinal distance
from $\Delta\varpi$ and multiplying by the $a,q$ weighting from above. In practice we grid our pole position distribution
as a {\it HEALPIX}\footnote{https://healpix.jpl.nasa.gov/html/idl.htm} map  (with NSIDE=32, for an approximately 1.8 degree resolution) and we calculate
separate pole position distributions for each value of 
$\Delta\varpi$ at one degree spacings. This three-dimensional
function is the probability that an {\it unbiased survey}
that found a KBO with $a_j$ and $q_j$ would have found that object
with $(i,\Delta \varpi,\Delta \Omega)_j$ in
the $k^{th}$ simulation, or
\begin{equation}
P_{j,k}^{(a,e)_{j}}[(i,\Delta\varpi,\Delta\Omega)_j|(m_9,a_9,e_9,i_9)_k].
\end{equation}
For arbitrary values of $\varpi_9$ and $\Omega_9$,
this probability distribution can be translated to an absolute frame of reference with simple rotations to give
\begin{equation}
P_{j,k}^{(a,e)_{j}}[(i,\varpi,\Omega)_j|(m_9,a_9,e_9,i_9)_k,\varpi_9,\Omega_9].
\end{equation}

\subsection{Likelihood}
With functions now specified for the probability of detecting
object $j$ at $(i,\varpi,\Omega)_j$ and also for the probability
of detecting object $j$ at $(i,\varpi,\Omega)_j$ assuming a
uniform distribution across the sky, we can calculate our
biased probability distribution for object $j$ in simulation $k$, 
$P'_{j,k}$, by simple multiplication:
\begin{multline}
P_{j,k}^{\prime (a,e)_{j}}[(i,\varpi,\Omega)_j|(m_9,a_9,e_9,i_9)_k,\varpi_9,\Omega_9]=\\
P_{j,k} \times B_j^{(a,e,H)_j}[(i,\varpi,\Omega)_j|U]
\end{multline}
where the arguments for $P_{j,k}$ are omitted for simplicity.
We rewrite this probability as our likelihood function
in the form of Equation (2) and take the product
of the individual $j$ likelihoods to form the overall
likelihood for each model $k$ at the values of
$\varpi_9$ and $\Omega_9$:
\begin{equation}
    \mathcal{L}_k[(m_9,a_9,e_9,i_9)_k,\varpi_9,\Omega_9|X],
\end{equation}
where $X$ represents the full set of orbital elements of the
distant KBOs from Table 2. The likelihood is discretely sampled by
the numerical models in the first four parameters and 
continuously sampled analytically in the two angular 
parameters.

The likelihoods sparsely sample a seven-dimensional,
highly-correlated parameter space. With even
a cursory examination of the likelihoods, however, 
several trends are apparent (Figures 5 and 6). First,
the model with the maximum likelihood, $M_9=5$ M$_{\rm earth}$, $a_9=300$ AU, $i_9=17^{\circ}$, $e_9=0.15$, $\varpi_9=254^{\circ}$, and $\Omega_9=108^{\circ}$,
is nearly a local peak in every dimension. Semimajor axes
inside of $\sim$300 AU lead to low likelihoods,
but more distant Planets Nine are viable
(particularly if they are more massive), even if at
reduced likelihood. The inclination appears quite 
well confined to regions near 15$^\circ$, and strong peaks
near $\varpi_9=250^{\circ}$ and $\Omega_9=100^{\circ}$ are evident.

\begin{figure*}
\epsscale{.9}
\plotone{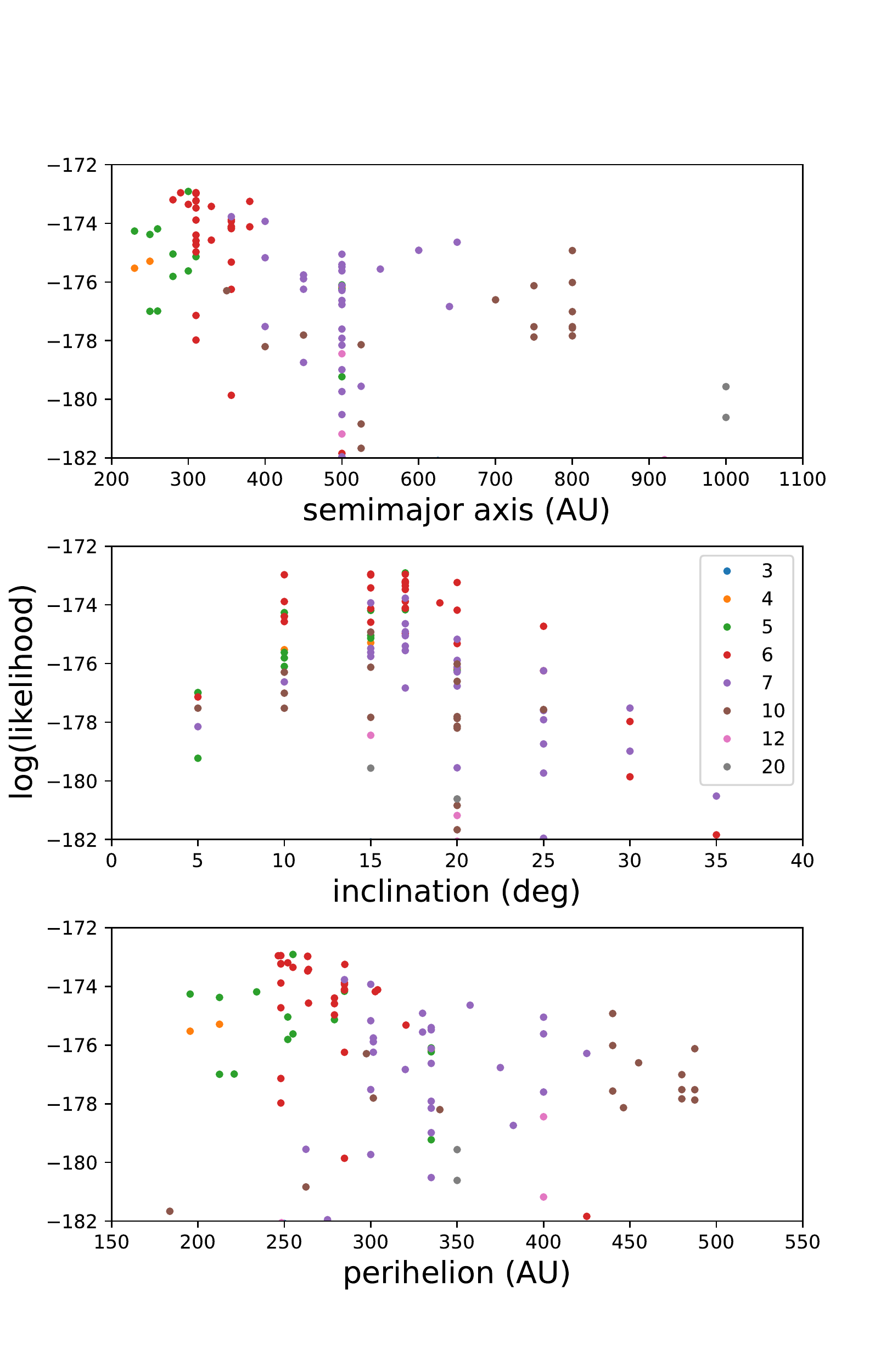}
    \caption{One dimensional plots of the log(likelihood) values as
    function of  semimajor axis ($a_9$), inclination ($i_9$)
    and perihelion distance ($q_9$), at the maximum likelihood
    in $\Omega_9$ and $\varpi_9$ for each simulation. The points are
    colored by the mass of Planet Nine ($m_9$) as shown in the legend. 
    The one-dimensional plots show the general behavior
    but do not show the significant
    correlations between parameters.}
    \label{fig:my_label}
\end{figure*}

\begin{figure}
\epsscale{1.3}
\plotone{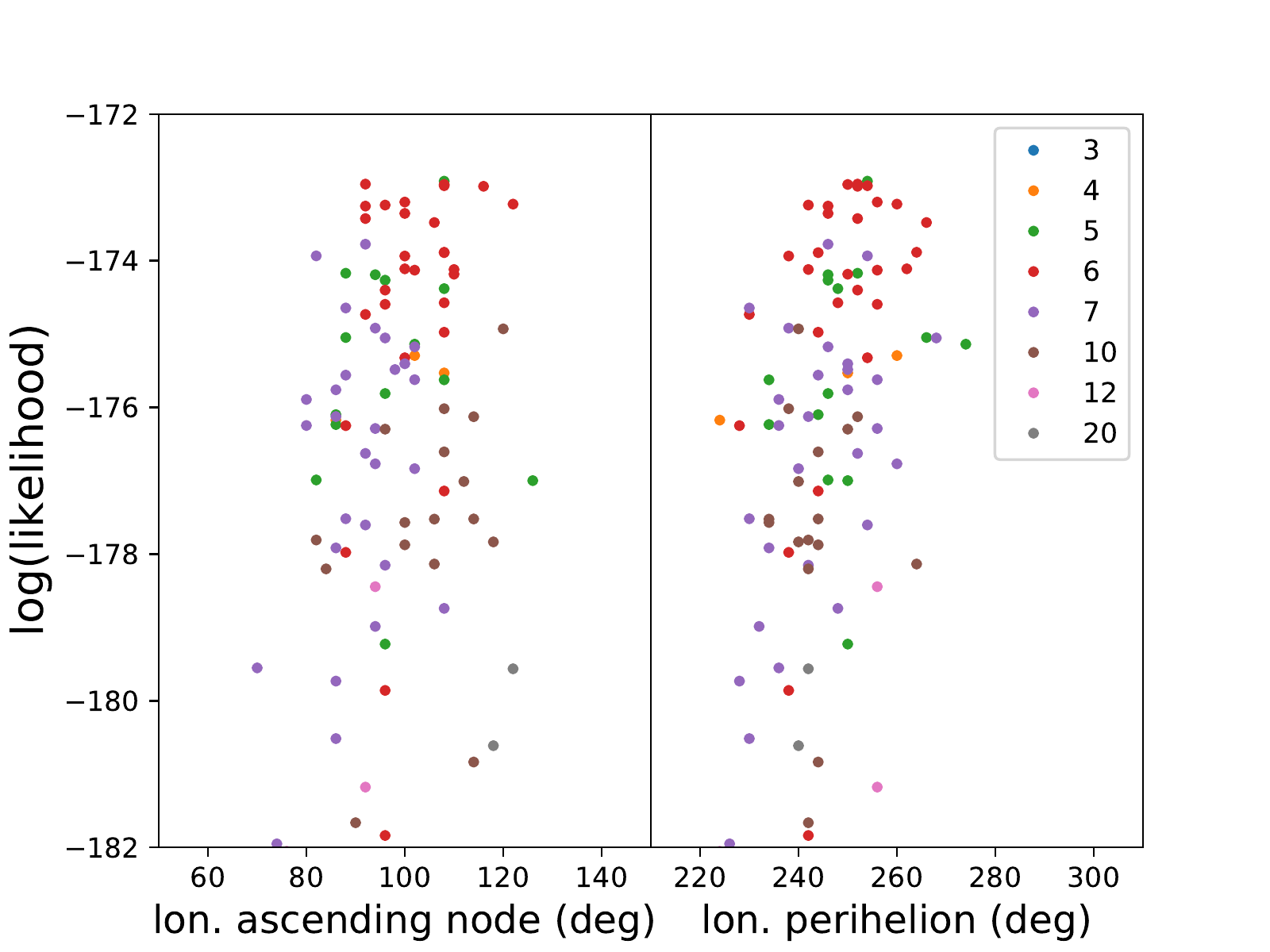}
\caption{The longitude of ascending node ($\Omega_9$)
and the longitude of perihelion ($\varpi_9$)
at which the maximum likelihood occurs in each of the simulations.
The points are colored by the mass of Planet Nine ($m_9$).}
\end{figure}

\subsection{Gaussian process emulation}
To further explore the orbital parameters, their 
correlations, and their uncertainties, 
we require a continuous, rather than discretely sampled, 
likelihood function.
To estimate this likelihood at an arbitrary value of 
$(m_9,a_9,i_9,e_9,\varpi_9,\Omega_9)$
we perform the following steps.
First, because the likelihoods as functions
of $\varpi_9$ and $\Omega_9$ are densely sampled for each
simulation, we perform a simple interpolation to obtain
an estimated likelihood for each simulation at the specific
desired value of $\varpi_9$ and $\Omega_9$.
We next take the 121 simulations with their now-interpolated
likelihoods and  use these to create a computationally
inexpensive Gaussian Process model as an emulator
for the likelihoods.
The behaviour of the likelihoods is extremely asymmetric, in 
particular in $m_9$ and $a_9$, with likelihood falling rapidly 
at small values of $m_9$ but dropping only slowly at higher
values. Likewise, the likelihoods change rapidly for small
values of $a_9$, while changing more slowly at higher $a_9$. 
To better represent this behaviour, we rescale the variables that
we use in our Gaussian Process modeling. We use $a'=(a_9/m_9)^{-0.5}$ and
we replace $e_9$ with a similarly-scaled function of perihelion distance,
$q'=\{a_9*(1-e_9)/m_9\} ^{-0.5}$. These scalings cause the likelihoods to
appear approximately symmetric about their peak values and to peak at
similar values of $a'$ and $q'$ for all masses (Figure 6). 
To enforce the smoothness and symmetry in
the Gaussian Process model, 
we choose a Mate\'rn kernel, which allows for a freely
adjustable smoothness parameter, $\nu$. We chose a value
of $\nu=1.5$, corresponding to a once-differentiable function,
and which appears to adequately reproduce the expected behavior
of our likelihood models. We force the length scales of the Mat\'ern
kernel to be within the bounds (0.5, 2.0), (0.02, 0.05), 
(1.0, 10.0), and (1.0,100.0) for
our 4 parameters and for units of earth masses, AU, and degrees, corresponding 
to the approximate correlation length scales that we see in the likelihood simulations. 
We multiply this kernel by a constant kernel and also add a constant kernel.
Beyond the domain of the simulations we add artificial points with low
likelihood to prevent unsupported extrapolation.
The model is implemented using scikit-learn in Python \citep{scikit-learn}.

The emulator produces a likelihood value at arbitrary values
of $(M_9,a_9,i_9,e_9,\varpi_9,\Omega_9)$, and appears to do a
reasonable job
of reproducing the likelihoods of the numerical models, interpolating
between these models, and smoothly extending the models over the
full range of interest. 
Figure 7 gives an example of the correspondence between 
individual measured likelihoods and the emulator in the rescaled variable
$a'$. Viewed in the rescaled variables, the likelihoods and the emulator are relatively 
regular, symmetric, and well-behaved. Similar results are seen for
$i_9$ and $q_9$. While the emulator does not perfectly
reproduce the simulation likelihoods, the large-scale behavior is captured
with sufficient fidelity to allow us to use these results for
interpolation between the discrete simulations.

\begin{figure*}
\epsscale{0.8}
\plotone{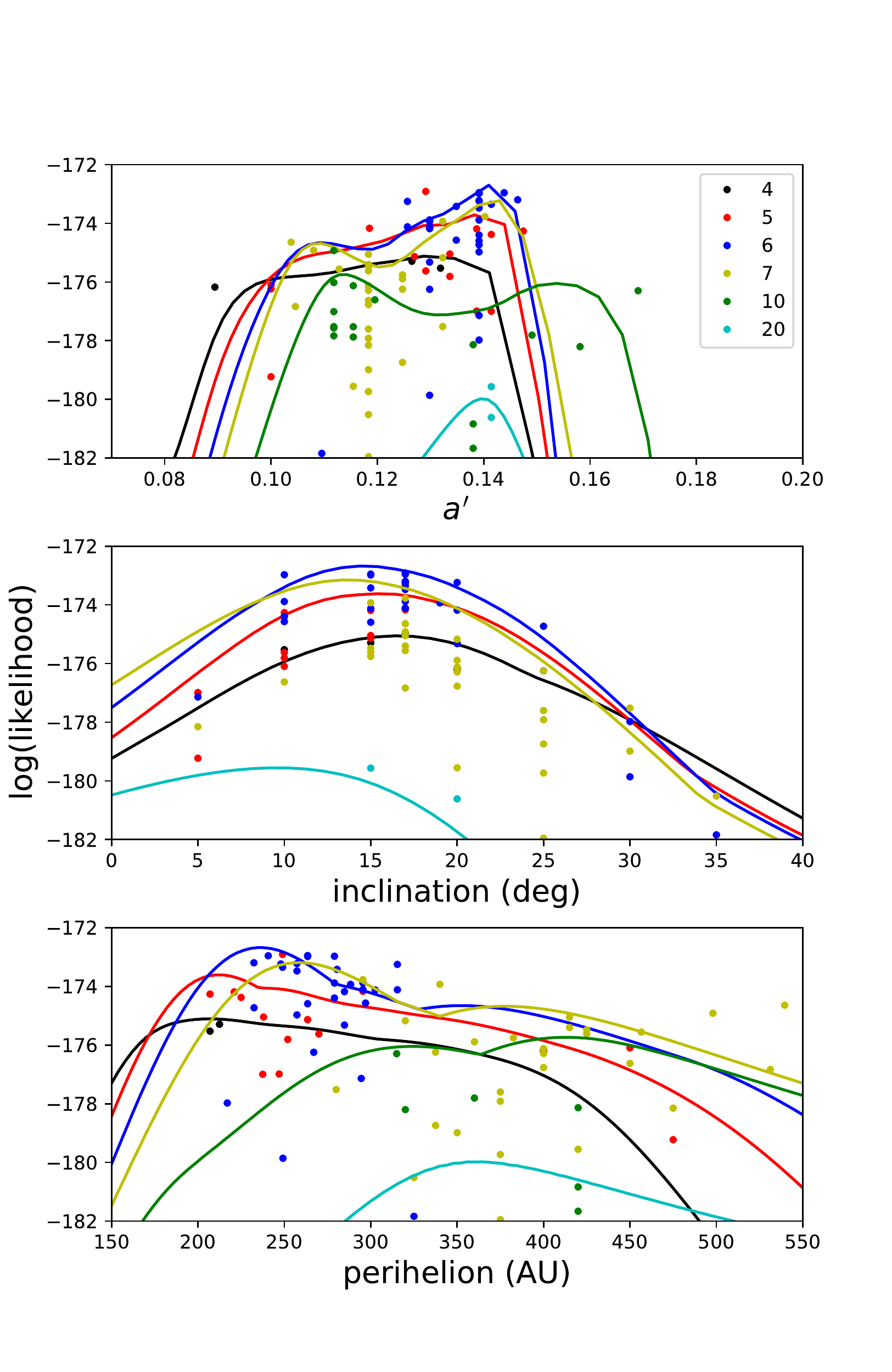}
\caption{The discrete log(likelihoods) from the simulation versus 
the continuous results from the emulator. 
Results from different values of $m_9$ are
shown as different colors and labeled in the legend. 
The top panel shows
the rescaled parameter $a'=(a_9/m_9)^{-0.5}$. 
In this rescaled
variable the likelihoods are approximately symmetric and 
vary smoothly as a function of mass different masses. The lines
show the output from the Gaussian Process emulator with a fixed
value of $i_9=15^\circ$, maximum likelihood values of $\Omega_9$ and
$\varpi_9$ and an iterative search to find the maximum likelihood
value for $e_9$. The emulator result shown thus represents the highest
possible maximum likelihood at $i_9=15^\circ$
for each value of $a'$ for each $m_9$.
The middle panel shows emulator results versus inclination
for the maximum likelihood values of all parameters, while the bottom
panel shows emulator results versus perihelion at a fixed inclination of
$i_9=15^\circ$ and maximum likelihood values of the other parameters.
In all cases, the emulator output follows the upper envelope of
the likelihoods, as expected. 
}
\end{figure*}

\subsection{MCMC}
We use this Gaussian Process emulator to produce a Markov Chain Monte Carlo (MCMC)
model of the mass and orbital parameters of Planet Nine. 
We use the Python package {\it emcee} \citep{2013_Foreman-Mackey} 
which implements 
the \citet{ISI:000282653600004} affine-invariant MCMC ensemble sampler.
We consider two different priors for the semimajor axis distribution.
The Planet Nine hypothesis is agnostic to a formation mechanism for
Planet Nine, thus a uniform prior in semimajor axis seems 
appropriate. Nonetheless, different formation mechanisms 
produce different semimajor axis distributions. Of the
Planet Nine formation mechanisms, ejection from the Jupiter-Saturn
region followed by cluster-induced perihelion raising is the most
consistent with known solar system constraints
\citep{2019PhR...805....1B}. 
In \citet{2021ApJ...910L..20B} we consider this process and find a 
distribution of expected
semimajor axes that smoothly rises from about 300 AU to
a peak at about 900 AU before slowly declining. The distribution from
these simulations can be empirically fit by a Fr\'echet distribution of
the form $p(a)=(a-\mu)^{-(\alpha+1)} \exp(-((a-\mu)/\beta)^{-\alpha})$ with
$\alpha = 1.2$,
$\beta = 1570$ AU, and
$\mu = -70$ AU. We consider both this and the uniform prior
and discuss both below.
Additionally,
we assume 
priors of $\sin(i_9)$ in inclination and $e_9$ in eccentricity to account 
for phase-space volume. Priors in the other parameters are uniform. 
We sample parameter space using 100 separate chains (``walkers'')
with which we obtain 20890 samples each.
We use the {\it emcee} package to  calculate 
the autocorrelation scales
of these chains and find that maximum is 130 steps, which is 160
times smaller than the length of the chain, ensuring that the
chains have converged.
We discard the initial 260 steps of each chain as burn-in and sample 
each every 42 steps to obtain 49100 uncorrelated samples.

Examining the two different choices of prior for $a_9$ we see
that the posterior distributions of the
angular parameters, $i_9$, $\Omega_9$, and $\varpi_9$,
are unchanged by this choice. The parameters $m_9$, $a_9$, and $e_9$
are, however, affected. This effect can best be seen 
in the posterior 
distributions of $a_9$ for the two different priors. 
The  uniform prior
has 16th, 50th, and 84th percentile values of
$a_9=300$, 380, and 520 AU ($380^{+140}_{-80}$ AU) 
versus 
$a_9=360$, 460, and 640 AU ($460_{-100}^{+180}$ AU)
for 
the cluster scattering prior. While the two posterior distributions
agree within $1 \sigma$, the differences are sufficiently
large that predictions of expected magnitude, for example,
could be affected.
Here we will retain the uniform 
prior for continued analysis, but we keep in mind below
the effects of a semimajor axis distribution with values
approximately 20\% larger.
For this uniform prior, the marginalized
perihelion and aphelion distances of Planet Nine
are $300^{+85}_{-60}$ and $460^{+200}_{-110}$ AU,
respectively. 

Figure 8 shows a corner plot illustrating the full 
two-dimensional correlation
between the posterior distribution of
pairs of parameters for the cluster scattering prior in $a_9$.
We see the clear expected correlations related to $a_9$, $m_9$, and $e_9$.
No strong covariances exist between the other parameters. The posterior
distributions for $i_9$ and $\Omega_9$ are among the most tightly
confined, suggesting that the data strongly confine the pole
position -- and thus orbital path through the sky -- of Planet Nine.
\begin{figure*}
\plotone{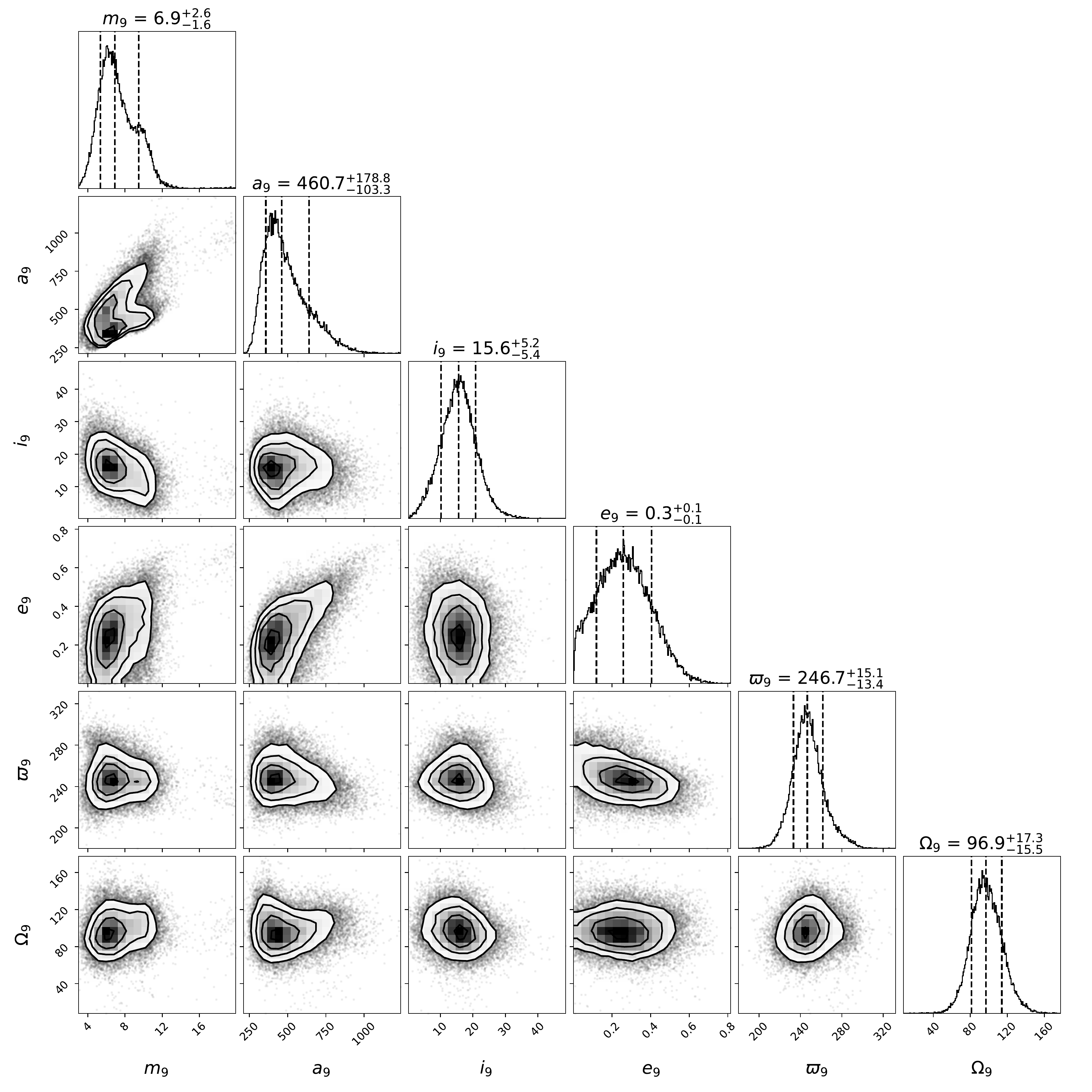}
\caption{A corner plot showing the histograms and covariances of
the parameters. The dashed lines on the histograms show
the median and 16th and 84th percentiles of each marginalized
distribution. The two dimensional histograms include
the 1, 2, and 3$\sigma$ contour lines.}
\end{figure*}

Examination of Fig. 1 helps to explain why low values
of $m_9$ and $a_9$ are preferred. The mass is directly
related to the width of the cluster, and masses
greater than 6 M$_{\rm earth}$ lead to narrower clusters
than those observed. Likewise, a low $m_9$ planet
requires a small semimajor axis to have a distance to
the wall of only $\sim$200 AU as the data appear to support.
It is possible, of course, that the two KBOs with $a\sim200$~AU are only coincidentally situated within
the cluster and the real wall, and thus $a_9$ is more
distant, but the likelihood analysis correctly 
accounts for this possibility.

\section{The predicted position and brightness of Planet Nine}
With distributions for the mass and orbital elements of Planet Nine
now estimated, we are capable of determining the probability distribution
of the on-sky location, the heliocentric distance,
and the predicted brightness of Planet Nine.
We first  use the full set of samples from the MCMC to determine the probability
distribution function of the sky position and heliocentric distance
of Planet Nine. To do so we
calculate the heliocentric position of an object with the orbital parameters
of each MCMC sample at one degrees spacings in mean anomaly, $M_9$. The sky
density of these positions is shown in Figure 9. Appropriately normalized,
this sky plane density represents the probability distribution function
of finding Planet Nine at any heliocentric position in the sky.
Approximately 95\% of the probability is within a swath of the sky
that is
$\pm12^\circ$ in declination from an orbit with an inclination of
$16\circ$ and an ascending node of$97^\circ$, the median marginalized
values of these parameters.
\begin{figure*}
    \includegraphics[scale=.85]{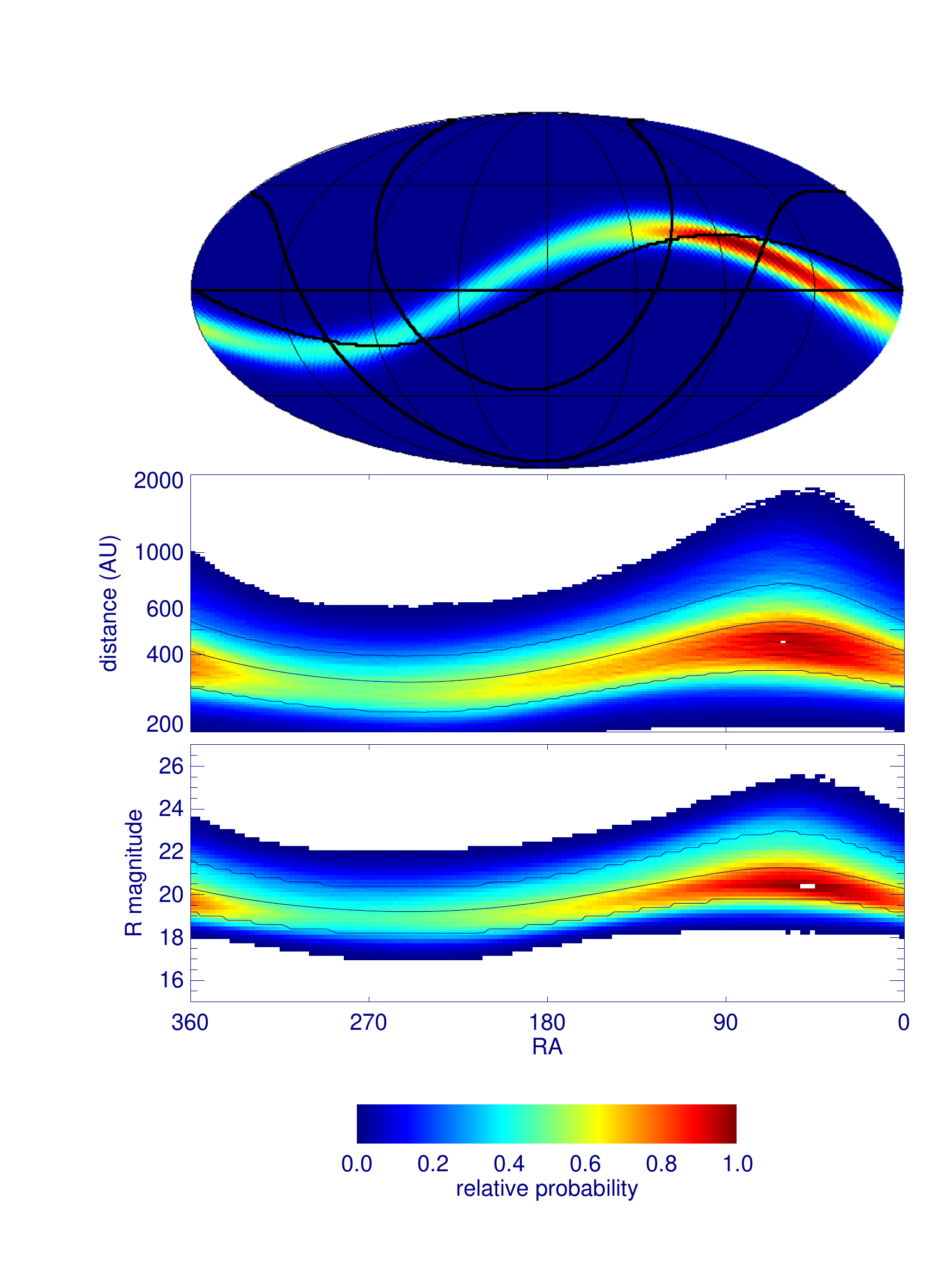}
    \caption{The sky plane density, heliocentric distance, and
    predicted $R$ magnitude for our Planet Nine samples. The 
    top panel is a Mollweide equal area projection centered
    at an RA of 180 and declination of 0. Thick lines show the
    celestial equator, the ecliptic, and $\pm20^\circ$ from the
    galactic plane. Thin lines show every 45$^\circ$ of RA and
    declination. The middle panel shows the probability distribution
    of heliocentric distance as a function of RA. The lines show
    the 16th, 50th, and 84th percentiles of the distribution. The 
    bottom panel shows modeled R-band magnitudes again with the
    16th, 50th, and 84th percentile distributions.The peak in
    probability density at R.A.$\sim 45$ corresponds to the 
    predicted aphelion position of P9 where an eccentric object spends
    more of its time.}
    
\end{figure*}

To estimate the magnitude of Planet Nine we need not just the mass,
but also the diameter and the albedo, neither of which we directly
constrain. We thus model what we consider to be
reasonable ranges for these parameters.

For masses between 4-20 M$_{\rm earth}$ we assume that the 
most likely planetary composition is that of a sub-Neptune,
composed of an icy-rocky core with a H-He rich envelope
(we discuss alternatives below).
We assume a simple mass-diameter relationship of 
$r_9=(m_9/{3 \rm M}_{\rm earth})$ R$_{\rm earth}$ based
on fits to (admittedly much warmer) planets in this
radius and mass range by \citet{2013ApJ...772...74W}. The albedo of such an object has been modeled by \citet{2016ApJ...824L..25F},
who find that all absorbers are condensed out of the
atmosphere and the planet should have a purely Rayleigh-scattering
albedo of $\sim$0.75. We conservatively assume a full range of albedos
from 0.2 -- half that of Neptune -- to 0.75. With
these diameters and albedos we can use the modeled distances to
determine the brightness of Planet Nine for each of the samples. 
Figure 8 shows the predicted magnitudes of Planet Nine. 
At the brightest end, Planet Nine could already have been
detected in multiple surveys, while at the faintest 
it will require dedicated searches on 8-10 meter telescopes.

\section{Caveats}
Both the maximum likelihood and the fully marginalized
MCMC posterior distributions suggest that Planet Nine
might be closer and potentially brighter than previously expected.
The original 
analysis of \citet{2016AJ....151...22B} was a simple
proof-of-concept that an inclined eccentric massive planet
could cause outer solar system clustering, so the
choice of $m_9$=10 $M_{\rm earth}$, $e_9$=0.7, and
$i_9=30^{\circ}$ was merely notional.
\citet{2016ApJ...824L..23B} showed that a wide range of
masses and semimajor axes were acceptable with the constraints
available at the time,
while \citet{2019PhR...805....1B} showed hints of a preference
for lower mass and semimajor axis.
As previously discussed, one of the strongest drivers for the lower
mass and semimajor axis of Planet Nine is the width of the longitude 
of perihelion cluster. With longitudes of perihelion ranging from 7 to 118$^\circ$, this 111$^\circ$ wide cluster is best matched by low masses, which
necessitates low semimajor axes to bring the wall in as close as 200 AU. 

One possibility for artificially widening the longitude of perihelion
cluster is contamination by objects
recently scattered into the $150<a<1000$ AU, $q>42$ AU region.
It is plausible that 2013FT28, the major outlier outside of the cluster,
is one such recently Neptune-scattered object. While integration
of the orbit of 2013FT28 shows that it is currently metastable, with a 
semimajor axis that diffuses on $\sim$Gyr timescales,
and while we attempted to exclude all recent Neptune-scattered objects
by requiring $q>42$ AU, we nonetheless note that within the 200 Myr of
our simulations $\sim$20\% of the
objects that start as typical scattering objects
with $30<q<36$ AU and $a<150 $AU have diffused to the $q>42$ AU, $a>150$ AU
region. These diffusing objects are broadly clustered around
$\Delta \varpi\sim 0\circ$ instead of around
$\Delta \varpi\sim 180^\circ$ like the stable cluster.
2013FT28 is such a strong outlier, however, that whether it is a
contaminant from this route or not, its presence has little affect on
our final retrieved orbital parameters. No Planet Nine simulations are
capable of bringing it into a region of high likelihood.

A more worrisome possibility for inflating the width of the longitude
of perihelion clustering is the scattering inward of objects
from the inner Oort cloud \citep{2021ApJ...910L..20B}.
As noted earlier, we have no
clear way to discriminate against these objects, and while the most distant
objects are more likely to have originated from this exterior source,
such objects can be pulled down to small semimajor axes, too.
We have no understanding of the potential magnitude -- if any -- of
this potential contaminating source,
so we assess the maximum magnitude of the effect by systematically
examining the exclusion of objects from the data set. Limiting the
number of objects under consideration will necessarily raise the
uncertainties in the extracted parameters, but we instead here simply look
at how it changes the maximum likelihood simulation.

We recalculate the maximum likelihood values of each simulation after exclusion of the 
object most distant from the average $\varpi$ 
position of the cluster (with the exception
of 2013FT29, which we always retain).
Even after excluding the 6 most extreme
objects in the cluster and retaining only
4, the maximum likelihood changes only from
$m_9=5$ to $m_9=6$ M$_{\earth}$ and from $a_9=300$ to
$a_9=310$ AU. The orbital angles do not change
substantially. 

We conclude that the preference for smaller values of mass and semimajor 
axis is robust, and that the orbital angles ($i_9, \Omega_9, \varpi_9$) are largely
unaffected by any contamination. While the posterior distributions
for $m_9$ and $a_9$ have large tails towards larger values, the possibility
of a closer brighter Planet Nine needs to be seriously considered.

An additional uncertainty worth considering is the diameter and albedo of
Planet Nine. We have assumed values appropriate for a gas-rich sub-Neptune
which, {\it a priori}, seems the most likely state for such a distant
body. Given our overall ignorance of the range of possibilities in the 
outer solar system, we cannot exclude the possibility of an icy body
resembling, for example, a super-Eris. Such an icy/rocky body
could be $\sim$50\% smaller than an equivalent
sub-Neptune in this mass range \citep{2014ApJ...792....1L}, 
and while the large KBOs 
like Eris have high
albedos, much of this elevated albedo could be driven by frost 
covering of darker irradiated materials as the objects move through
very different temperature regimes on very eccentric orbits. An
object at the distance of Planet Nine -- which stays below
the condensation temperature of most volatiles at all times --
could well lack such volatile
recycling and could have an albedo closer to the $\sim$10\% of
the large but not volatile-covered KBOs \citep{2008ssbn.book..335B}. 
Overall the effect of a smaller diameter and smaller albedo could make 
Planet Nine $\sim$ 3 magnitudes dimmer. Such a situation would make
the search for Planet Nine considerably more difficult. While 
the possibility of a dark super-Eris Planet Nine seems unlikely, it cannot be excluded.

Finally, we recall the affect of the choice of the prior on $a_9$. 
A prior assuming formation in a cluster
would put Planet Nine more distant than shown here,
though it would also predict higher masses. Combining those 
effects we find that the magnitude distribution seen in Figure 
8 would shift fainter by about a magnitude near aphelion but would
change little close to perihelion.

While all of these caveats affect the distance, mass, and brightness
of Planet Nine, they have no affect on the sky plane position
shown in Figure 8. To a high level of confidence, Planet Nine
should be found along this delineated path.

\section{Conclusion}
We have presented the first estimate of Planet Nine's mass and orbital
elements using a full statistical treatment of the likelihood of
detection of the 11 objects with $150<a<1000$ AU and $q>42$ AU as
well as the observational biases associated with these detections. 
We find that the median expected Planet Nine semimajor axis is
significantly closer than previously understood, though the range of
potential distances remains large. At its brightest predicted 
magnitude, Planet Nine could well be in range of the large number of
sky surveys being performed with modest telescope, so we expect that
the current lack of detection suggests that it is not as the brightest
end of the distribution, though few detailed analysis of these surveys 
has yet been published. 

Much of the predicted magnitude range of Planet Nine is within the
single-image detection limit of the LSST survey of the Vera Rubin telescope,
$r\sim 24.3$, though the current survey plan does not extend as
far north as the full predicted path of Planet Nine.
On the faint end of the distribution, or if Planet Nine is unexpectedly
small and dark, detection will still require imaging with 10-m class
telescopes or larger. 

Despite recent discussions, statistical evidence for clustering in the outer
solar system remains strong, and a massive planet on a distant inclined
eccentric orbit remains the simplest hypothesis. Detection of
Planet Nine will usher in a new understanding of the outermost part
of our solar system and allow detailed study of a fifth giant planet
with mass common throughout the galaxy.

\acknowledgments This manuscript owes a substantial debt
to the participants at the 
\textit{MATH + X Symposium on Inverse Problems and Deep Learning in
Space Exploration} held at Rice University in Jan 2019 
with whom
we discussed the issue of inverting the observations of
KBOs to solve for Planet Nine. We would also like to thank
two anonymous reviewers of a previous paper whose excellent 
suggestions ended up being incorporated into this paper
and  \href{https://twitter.com/Snippy_X}{@Snippy\_X} and \href{https://twitter.com/siwelwerd}{@siwelwerd} on Twitter for advice on notation for 
our likelihood functions.

\software{HEALPix \citep{Gorski_2005},
astropy \citep{2013A&A...558A..33A},  
scikit-learn \citep{scikit-learn},
emcee \citep{2013_Foreman-Mackey},
corner \citep{corner}}
\
\eject
\
\eject

\startlongtable
\begin{deluxetable*}{crrrrrrrr}
\tablehead{
\colhead{$m_9$} & \colhead{$a_9$} & \colhead{$i_9$} & \colhead{$e_9$} & \colhead{$\varpi_9$} & \colhead{$\Omega_9$} & \colhead{$\ell$} & \colhead{$\Delta \ell$} & \colhead{num.} \\
\colhead{($M_{\rm earth}$)} & \colhead{(AU)} & \colhead{(deg)} & & \colhead{(deg)} & \colhead{(deg)} & & & \colhead{particles}
}
\startdata
 3& 625&15&0.60&356&166&-182.1& -9.2& 21100\\
 4& 230&10&0.15&250&108&-175.5& -2.6& 30000\\
 4& 250&15&0.15&260&102&-175.3& -2.4& 30000\\
 4& 500&20&0.33&224& 86&-176.2& -3.3&120500\\
 5& 230&10&0.15&246& 96&-174.3& -1.4& 30000\\
 5& 250& 5&0.15&250&126&-177.0& -4.1& 30000\\
 5& 250&10&0.15&248&108&-174.4& -1.5& 30000\\
 5& 260&15&0.10&246& 94&-174.2& -1.3& 25600\\
 5& 260& 5&0.15&246& 82&-177.0& -4.1& 30000\\
 5& 280&10&0.10&246& 96&-175.8& -2.9& 25600\\
 5& 280&15&0.10&266& 88&-175.0& -2.1& 25600\\
 5& 300&10&0.15&234&108&-175.6& -2.7& 25600\\
 5& 300&17&0.15&254&108&-172.9&  0.0& 25600\\
 5& 310&15&0.10&274&102&-175.1& -2.2& 25600\\
 5& 356&17&0.20&252& 88&-174.2& -1.3& 25600\\
 5& 500& 5&0.33&250& 96&-179.2& -6.3& 25600\\
 5& 500&10&0.33&244& 86&-176.1& -3.2& 25500\\
 5& 500&20&0.33&234& 86&-176.2& -3.3& 20200\\
 5& 720&20&0.65&234& 96&-185.1&-12.2& 30100\\
 6& 280&17&0.10&256&100&-173.2& -0.3& 25500\\
 6& 290&17&0.15&250&108&-173.0& -0.0& 25600\\
 6& 300&17&0.15&246&100&-173.4& -0.4& 25600\\
 6& 310&10&0.10&252& 96&-174.4& -1.5& 25600\\
 6& 310&15&0.10&256& 96&-174.6& -1.7& 25600\\
 6& 310&17&0.10&244&108&-175.0& -2.1& 25600\\
 6& 310&10&0.15&256&108&-173.0& -0.1& 25600\\
 6& 310&15&0.15&252&116&-173.0& -0.1& 25600\\
 6& 310&17&0.15&266&106&-173.5& -0.6& 19900\\
 6& 310& 5&0.20&244&108&-177.1& -4.2& 25600\\
 6& 310&10&0.20&244&108&-173.9& -1.0& 25000\\
 6& 310&15&0.20&252& 92&-173.0& -0.0& 25400\\
 6& 310&17&0.20&260&122&-173.2& -0.3& 13600\\
 6& 310&20&0.20&242& 96&-173.2& -0.3& 23700\\
 6& 310&25&0.20&230& 92&-174.7& -1.8& 20000\\
 6& 310&30&0.20&238& 88&-178.0& -5.1& 25500\\
 6& 330&10&0.20&248&108&-174.6& -1.7& 31300\\
 6& 330&15&0.20&252& 92&-173.4& -0.5& 14400\\
 6& 356&20&0.10&254&100&-175.3& -2.4& 25600\\
 6& 356&20&0.15&250&110&-174.2& -1.3& 25600\\
 6& 356&15&0.20&256&102&-174.1& -1.2& 21200\\
 6& 356&17&0.20&262&100&-174.1& -1.2& 25600\\
 6& 356&17&0.20&264&108&-173.9& -1.0& 25600\\
 6& 356&19&0.20&238&100&-173.9& -1.0& 48500\\
 6& 356&25&0.20&228& 88&-176.2& -3.3& 40200\\
 6& 356&30&0.20&238& 96&-179.9& -6.9& 16700\\
 6& 380&17&0.20&242&110&-174.1& -1.2& 25600\\
 6& 380&17&0.25&246& 92&-173.3& -0.3& 25600\\
 6& 500&35&0.15&242& 96&-181.8& -8.9& 30000\\
 6& 600&40&0.15&260& 94&-184.0&-11.1& 30000\\
 6& 800&50&0.15&242& 82&-188.4&-15.5& 30000\\
 7& 356&17&0.20&246& 92&-173.8& -0.9& 25600\\
 7& 400&15&0.25&254& 82&-173.9& -1.0& 30900\\
 7& 400&20&0.25&246&102&-175.2& -2.3& 52800\\
 7& 400&30&0.25&230& 88&-177.5& -4.6& 30800\\
 7& 450&25&0.15&248&108&-178.7& -5.8& 30000\\
 7& 450&15&0.33&250& 86&-175.8& -2.8& 29700\\
 7& 450&20&0.33&236& 80&-175.9& -3.0& 25600\\
 7& 450&25&0.33&236& 80&-176.2& -3.3& 23500\\
 7& 500&20&0.15&256& 94&-176.3& -3.4& 25600\\
 7& 500&15&0.20&256&102&-175.6& -2.7& 25600\\
 7& 500&17&0.20&268& 96&-175.1& -2.1& 25600\\
 7& 500&25&0.20&254& 92&-177.6& -4.7& 25600\\
 7& 500&20&0.25&260& 94&-176.8& -3.9& 25600\\
 7& 500& 5&0.33&242& 96&-178.2& -5.2& 57300\\
 7& 500&10&0.33&252& 92&-176.6& -3.7& 41400\\
 7& 500&15&0.33&250& 98&-175.5& -2.6& 47700\\
 7& 500&17&0.33&250&100&-175.4& -2.5& 17500\\
 7& 500&20&0.33&242& 86&-176.1& -3.2& 52400\\
 7& 500&25&0.33&234& 86&-177.9& -5.0& 54000\\
 7& 500&30&0.33&232& 94&-179.0& -6.1& 59600\\
 7& 500&35&0.33&230& 86&-180.5& -7.6& 41700\\
 7& 500&25&0.40&228& 86&-179.7& -6.8& 35000\\
 7& 500&25&0.45&226& 74&-182.0& -9.0& 27700\\
 7& 525&20&0.50&236& 70&-179.6& -6.6& 33000\\
 7& 550&17&0.40&244& 88&-175.6& -2.6& 25600\\
 7& 600&17&0.45&238& 94&-174.9& -2.0& 25600\\
 7& 640&17&0.50&240&102&-176.8& -3.9& 16900\\
 7& 650&17&0.45&230& 88&-174.6& -1.7& 25500\\
 7& 800&50&0.15&310& 50&-190.4&-17.5& 30000\\
 7& 830&20&0.70&208& 96&-184.7&-11.7& 51200\\
 7&1000&60&0.15&298& 94&-191.2&-18.3& 30000\\
 8& 400&20&0.15&248&108&-177.1& -4.2& 30000\\
10& 350&10&0.15&250& 96&-176.3& -3.4& 30000\\
10& 400&20&0.15&242& 84&-178.2& -5.3& 30000\\
10& 450&20&0.33&242& 82&-177.8& -4.9& 34300\\
10& 525&20&0.15&264&106&-178.1& -5.2& 30000\\
10& 525&30&0.15&266&102&-184.6&-11.7& 30000\\
10& 525&40&0.15&304&138&-189.9&-17.0& 30000\\
10& 525&20&0.50&244&114&-180.8& -7.9& 39700\\
10& 525&20&0.65&242& 90&-181.7& -8.8& 20900\\
10& 525&30&0.65&244& 36&-187.1&-14.2& 35600\\
10& 700&20&0.35&244&108&-176.6& -3.7& 25600\\
10& 700&30&0.70&290&132&-190.0&-17.1& 25600\\
10& 750&10&0.35&234&106&-177.5& -4.6& 19500\\
10& 750&15&0.35&252&114&-176.1& -3.2& 22400\\
10& 750&20&0.35&244&100&-177.9& -5.0& 25500\\
10& 800& 5&0.40&244&114&-177.5& -4.6& 25600\\
10& 800&10&0.40&240&112&-177.0& -4.1& 25600\\
10& 800&15&0.40&240&118&-177.8& -4.9& 25600\\
10& 800&15&0.45&240&120&-174.9& -2.0& 25600\\
10& 800&20&0.45&238&108&-176.0& -3.1& 28600\\
10& 800&25&0.45&234&100&-177.6& -4.7& 23500\\
10& 800&30&0.45&242& 50&-184.0&-11.1& 16800\\
10& 800&60&0.45&182&114&-183.0&-10.1& 30400\\
10& 870&20&0.73&254& 92&-185.4&-12.5& 17900\\
10&1000&60&0.15&314& 96&-192.8&-19.9& 23600\\
10&1400&70&0.15&224& 30&-190.0&-17.1& 30000\\
12& 500&15&0.20&256& 94&-178.4& -5.5& 25600\\
12& 500&20&0.20&256& 92&-181.2& -8.3& 25600\\
12& 500&25&0.20&266&102&-182.9&-10.0& 25600\\
12& 920&20&0.73&224& 76&-182.1& -9.1& 25800\\
12& 960&20&0.79&242& 54&-186.8&-13.9& 24900\\
14& 960&20&0.74&220& 76&-185.6&-12.7& 28000\\
16&1000&20&0.75&248& 76&-183.2&-10.2& 33600\\
20& 900&60&0.15&306& 66&-189.0&-16.1& 30000\\
20&1000&15&0.65&242&122&-179.6& -6.7& 30100\\
20&1000&20&0.65&240&118&-180.6& -7.7& 33000\\
20&1000&25&0.65&246& 70&-185.5&-12.6& 32300\\
20&1070&20&0.77&240&124&-185.2&-12.3& 64900\\
20&1400&70&0.15&264&  0&-186.8&-13.9& 30000\\
20&2000&80&0.15&260&152&-190.1&-17.2& 30000\\
\enddata
\tablecomments{Parameters used in the numerical simulations on
the effects of Planet Nine ($m_9$, $a_9$, $i_9$, $e_9$) and 
the maximum ln(likelihood), $\ell$, which occurs at the 
listed value of $\varpi_9$ and $\Omega_9$. $\Delta \ell$ gives
the difference in ln(likelihood) from the maximum value, which
occurs at $m_9=5$, $a_9=310$, $i_9=15$, and $e_9=0.10$.}
\end{deluxetable*}


\end{document}